# Mechanisms behind high $CO_2$/$CH_4$ selectivity using ZIF-8 metal organic frameworks with encapsulated ionic liquids: a computational study

Tianhao Yu,[a,b] Qiong Cai,*[a] Guoping Lian,[a,c] Yinge Bai,[b] Xiaochun Zhang,[b] Xiangping Zhang,[b] and Lei Liu*[b]

[a] Department of Chemical Process Engineering, University of Surrey, Guildford, Surrey GU2 7XH, United Kingdom

[b] Beijing Key Laboratory of Ionic Liquids Clean Process, Institute of Process Engineering, Chinese Academy of Sciences, Beijing 100049, China

[c] Unilever Research Colworth, Colworth Park, Sharnbrook, Bedfordshire MK44 1LQ, United Kingdom

* Corresponding authors: liulei@ipe.ac.cn; q.cai@surrey.ac.uk

**Abstract**

$CO_2/CH_4$ separation using ionic liquids (ILs) encapsulated metal-organic frameworks (MOFs), especially ZIF-8, has shown promise as a new technique for separating $CO_2$ from $CH_4$. However, the mechanisms behind the high $CO_2/CH_4$ selectivity of the method remains indistinct. Here we report the progress of understanding the mechanisms from examining the ZIF-8 aperture configuration variation using DFT and MD simulations. The results indicate that the pristine aperture configuration exhibits the best separation performance, and the addition of ILs prevents the apertures from large swing (i.e. configuration variation). Subsequently, the effect of IL viscosity on the layout variation was investigated. MD simulations also show that the pristine aperture configuration is more stabilized by ILs with large viscosity (0-87Cp). Further increase of IL viscosity above 87Cp did not result in noticeable changes in the aperture stability.

# 1. Introduction

Natural gas is advantageous to oil and coal with low carbon emission.[1] $CO_2$ can comprise up to 70%[2] volume in natural gases, causing a series of potential problems including reduced heating values and gas qualities, corrosions of transport equipment, and high energy consumption in transport process. Up to date, several $CO_2$ separation technologies have been reported, e.g. pressure swing adsorption (PSA)[3–5], physical or chemical solvent scrubbing[6–10], and membrane separation[11–14]. In the PSA process, zeolites are good candidate adsorbents but their hydrophilicity has limited their application due to the presence of water vapour which reduces the available active surface area.[15] Amine, caustic solvent and amino acid salts are the most applied solvents in solvent scrubbing with amine scrubbing being the most mature technique.[6] Membrane-based techniques have been reported in the past few decades, and several types have been developed according to their nature (synthetic or biological), structure (symmetric or asymmetric), geometry (tubular, flat sheet or hollow fibre) or transport mechanism (dense or porous).[16] Though some separation techniques have shown promising separation effect, most of the them still suffer from high energy consumption and high cost,[17] thus more environment-friendly and economically viable $CO_2$ separation alternatives are desired. Materials with microscopic pores are expected to have superior separation performance due to their size exclusion mechanisms. Of particular interest are the so-called metal-organic frameworks (MOFs), which are formed by metal and organic linkers. Among this class of materials, ZIF-8 (ZIF: Zeolitic imidazolate framework) is one of the most investigated for gas separations due to its superior thermal and chemical stability.[18] It is formed by linking imidazolate and zinc atoms with a sodalite (SOD) topology (see Fig. 1). An SOD cage (diameter 1.12 nm) is connected to eight identical apertures (diameter 0.34 nm), formed by bridging six identical imidazolates. Each SOD cage (the primary unit cell) is then duplicated and generates three-dimensional supercells.

Ionic liquids (ILs) are salts comprised of organic cation and organic/inorganic ion pairs at liquid state. The most reported ILs are room-temperature ionic liquids (RTILs) which are at liquid state under room temperature. For the last decade, ILs (especially RTILs) have attracted great interest due to their novel properties such as high ionic conductivity, negligible vapour pressure, low flammability and electrochemical stability which make ILs potential candidates for $CO_2$ separation.[19] Extensive number of research papers have been reported since Blanchard *et al.*[20] first pointed out that $CO_2$ can be efficiently dissolved in ILs at room temperature and up to 40 MPa while ILs being insoluble in $CO_2$. Though pure ILs show a promising $CO_2$ separation performance, the technique suffers from high IL viscosity values. Some other promising IL-based methods such as IL-based solvents[21] and IL-based membranes[22,23] have also been tested for $CO_2$ separation but only at lab scale.

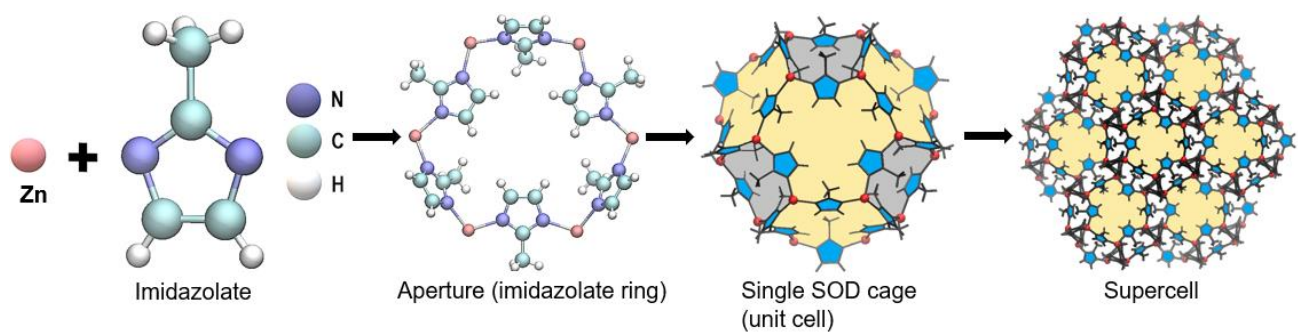

Figure 1. Schematic representation of the structure and topology of ZIF-8.

Theoretically, the apertures of ZIF-8 are expected to exhibit molecular sieving properties of $CO_2$ (0.33 nm) over larger molecules, such as $CH_4$ (0.38 nm) and $N_2$ (0.364 nm). However, experimental results[24,25] showed that molecules with size larger than the ZIF-8 framework aperture could still be adsorbed. In other words, the ZIF-8 shows a moderate $CO_2$ selectivity over $CH_4$ or $N_2$.[26] Research on improving ZIF-8 based $CO_2$ selectivity has been enlightened by a new approach reported by Ban *et al.*[27] The approach applies cage modification with cavity occupants by encapsulating the imidazolium-based [Bmim][Tf$_2$N] ionic liquids (ILs) molecules into the SOD cages, namely IL@ZIF-8 (as shown in Fig. 2(a)). Experimental results[27] show that such cage modification enhances the $CO_2/CH_4$ ideal selectivity from 7.5 to 41. Subsequently, much research efforts have been aroused in expanding such cage modification technique. For example, Koyuturk *et al.*[28] carried out studies on incorporating [Bmim][BF$_4$] in ZIF-8, and Zeeshan *et al.*[29] reported promising $CO_2/CH_4$ selectivity by confining [Bmim][SCN] into ZIF-8. On the other hand, several theoretical studies have been carried to gain the mechanistic profile of the cooperative effects of ILs and ZIF-8. Kinik *et al.*[28] studied the effect of encapsulating [Bmim][PF$_6$] into ZIF-8 with the aid of density functional theory (DFT) and Monte Carlo (MC) calculations. The authors concluded that the $CO_2$ selectivity was enhanced due to the new IL-created adsorption sites, which are occupied mostly by $CO_2$ molecules. Mohamed *et al.*[30] performed Grand Canonical Monte Carlo (GCMC) simulations to reproduce the experimental results reported by Ban *et al.*[27]. Moreover, Mohamed *et al.*[31] investigated the effect of distribution, composition and types of IL pairs on $CO_2/CH_4$ selectivity, and Thomas *et al.*[32] studied the effect of anion types on IL@ZIF-8 selectivity by DFT and GCMC simulations. The authors found that the hydrophobic fluorinated anions ([BF$_4$]$^-$, [Tf$_2$N]$^-$ and [PF$_6$]$^-$) gives a much better selectivity than hydrophilic non-fluorinated anions.

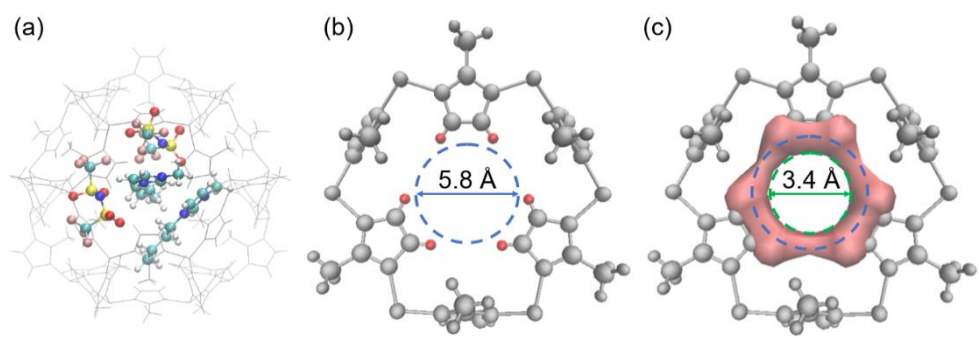

Figure 2. (a) Schematic representation of two [Bmim][Tf$_2$N] molecules encapsulated in ZIF-8 unit cell (IL@ZIF-8); Aperture diameter of ZIF-8: (b) the geometrical circumference (blue dashed circle) and (c) the effective ZIF-8 diameter (green dashed circle) by subtraction of van der Waals radius of hydrogen atom.

It is suggested that the presence of IL molecules in MOF cages narrows the effective $CO_2/CH_4$ passage, resulting in tailored molecular sieves[27,33]. Some other studies attributed the selectivity enhancement to IL anion types[32] and generated adsorption sites[28]. It appears that the detailed mechanisms regarding aperture configuration variation still remains indistinct. The aperture configuration refers to the specific shape of apertures. Due to the swing effect of ZIF-8, the imidazolate rings may rotate and Zn-imidazolate-Zn angles may vary from the pristine aperture, resulting in different aperture configurations. Therefore, the aperture configurations may have non-negligible effects on the gas mixture selection. For this reason, we perform DFT calculations and molecular dynamics (MD) simulations to investigate the effect of ZIF-8 aperture size changes and the effect on the $CO_2/CH_4$ selectivity. In DFT calculations, the energy barriers for $CO_2/CH_4$ molecules penetrating through the aperture are determined, where their deviation implies the aperture selectivity. In MD simulations, ZIF-8 and IL@ZIF-8 systems are simulated to investigate the influence of adding ILs ([$C_4$mim][$Tf_2$N]) on the changes of aperture size. Lastly, several ILs with different viscosities, such as [$C_2$mim][$Tf_2$N], [$C_6$mim][$Tf_2$N], [$C_8$mim][$Tf_2$N], [$C_2$mim][$PF_6$], [$C_2$mim][TFO] and [$C_2$mim][$BF_4$], are also simulated to study the effect of viscosity on the aperture size variations, and the selectivity of the IL@ZIF-8.

## 2. Methodology

In DFT calculations, $CO_2/CH_4$ molecules were placed 5 Å away from the aperture of ZIF-8 and moved closer to and eventually penetrated through the centre of the aperture, which is denoted as the centre of mass (COM) of the six Zn atoms. Fig. 3(a) illustrates an ideal process of $CO_2$ molecules diffusing through the ZIF-8 aperture: the $CO_2$ molecule was kept 'parallel' to the aperture, and the aperture shape remained unchanged. However, in realistic diffusing process, the diffusing molecule is not fixed at only one orientation and the aperture can oscillate due to the reported swing effect.[34] As it was not possible to cover all molecule orientations/aperture configuration variations in DFT calculation, we chose representative molecule orientations (given in Fig. 3(b)) and aperture configuration variations (given in Fig. 3(c)). In Fig. 3(b), two different orientations of $CO_2$ molecules were studied, denoted as parallel and vertical to the aperture, respectively. For $CH_4$ molecules, they were placed at orientations where vertex or triangle surface faces towards the aperture. Five different aperture configurations with altered sizes (given in Fig. 3(c)) were applied in the DFT calculations, namely pristine, *closed (d=4.1Å)*, *semi-open*, *closed (d=4.7Å)* and *open*. The configurations are different in the positions of the imidazolate rings and results in altered effective aperture sizes. Fig. 2(b) and (c) explains the calculation of the effective size of the aperture (green dashed circle), which is determined by subtracting the hydrogen van der Waals radius from the geometrical circumference (blue dashed circle). The geometrical circumference is the smallest circle out of all possible circles formed by arbitrary hydrogen atoms (coloured in red). It should be noted that the two *'closed'* configuration differs in the positions of the imidazolate rings, with the $CH_3$ group facing inwards and outwards respectively, giving rise to different aperture sizes, namely, *'closed (d=4.1Å)'*, and *'closed (d=4.7Å)'*. A recent study by Guo *et al.*[33] demonstrated possible existence of large size apertures, which were achieved by dilation through heat treatment to allow IL molecules to pass through. The aperture can be narrowed to its original size (3.4Å) upon cooling down to the room temperature, to ensure the successful confinement of ILs within ZIF-8 cages. In our DFT calculation, single point (SP) energies were calculated every 0.05Å (e.g. 5.00, 4.95, 4.90, … 0.15, 0.10,

0.05, 0.00 Å) when $CO_2$/$CH_4$ molecules (initially at 5.0 Å) were moving towards the ZIF-8 aperture (at 0.0 Å), by employing Gaussian 16[35] at B3LYP/6-311++G(d,p)[36] level of theory. The SP energy values were then plotted together to construct the potential energy curves.

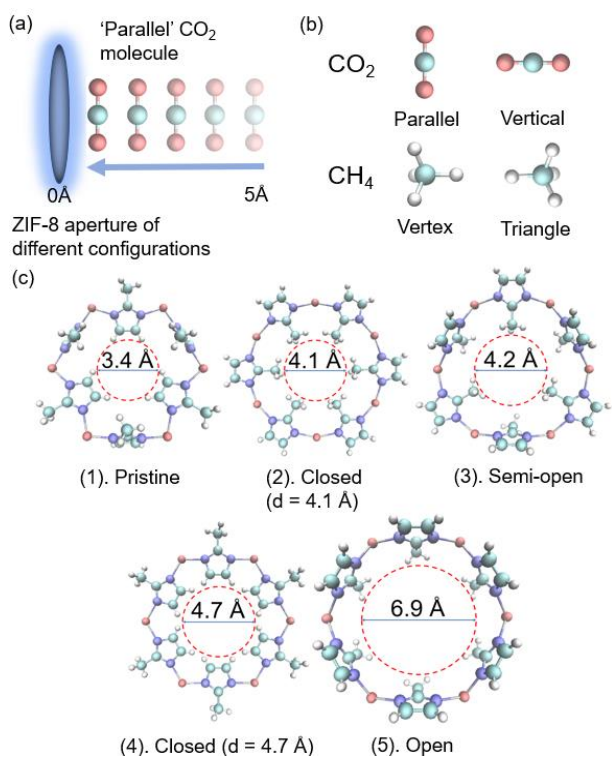

Figure 3. Schematic representation of DFT calculations $CO_2$/$CH_4$ energy barrier across ZIF-8 apertures. (a) $CO_2$/$CH_4$ moving through a ZIF-8 aperture. (b) two different orientations of $CO_2$ and $CH_4$ molecules. (c) Five different aperture structures with different aperture diameters.

MD simulations were performed for both ZIF-8 and IL@ZIF-8 unit cells applying the OPLS-AA[37] force field. GROMACS 2020.1[38] was used for all MD simulations, where the force field parameters for ZIF-8 and ionic liquids were taken from Wu *et al.*[39] and Doherty *et al.*[40] The ZIF-8 unit cell was constructed from the experimental ZIF document[18] using VESTA[41] and neutralised by removing excess imidazolates. Two IL pairs were randomly generated within the ZIF-8 cell cage using PACKMOL[42] to obtain the initial structure for IL@ZIF-8 for the first simulation. The IL@ZIF-8 structure was generated by packing further 398 IL pairs in a 6 nm × 6 nm × 6 nm cubic box to build a system of one ZIF-8 unit cell and 400 IL pairs. The IL pairs of different viscosity values including [$C_2$mim][$Tf_2$N], [$C_4$mim][$Tf_2$N], [$C_6$mim][$Tf_2$N], [$C_8$mim][$Tf_2$N], [$C_2$mim][$PF_6$], [$C_2$mim][TFO] and [$C_2$mim][$BF_4$] were changed for each simulation to investigate the effect of viscosity on aperture size variations. For each simulation, the total energy of the system was successfully minimised using the steepest descent method with convergence criteria $F_{max}<1000$ kJ·mol$^{-1}$·nm$^{-1}$ followed by an NVT equilibration of 1 ns. Periodic boundary conditions were applied in all three dimensions and Particle-Mesh-Ewald[43] summations were used to handle long-range interactions. The leapfrog algorithm[44] was used for integrating the equations of motion with a timestep of 1 fs. The temperature was kept constant at 298K using velocity rescaling with a stochastic term (v-rescale)[45]. After equilibration of temperature, an NPT ensemble was performed for 10 ns with pressure maintained at 1.0 bar using the Berendsen barostat[46]. All bonds were constrained using the LINCS algorithm[47] and the short-range electrostatic cut-off length was set to 14 Å. The production runs were performed for 100 ns after NPT equilibrium.

## 3. Results and Discussion

### 3.1 Effects of aperture configuration on $CO_2/CH_4$ separation

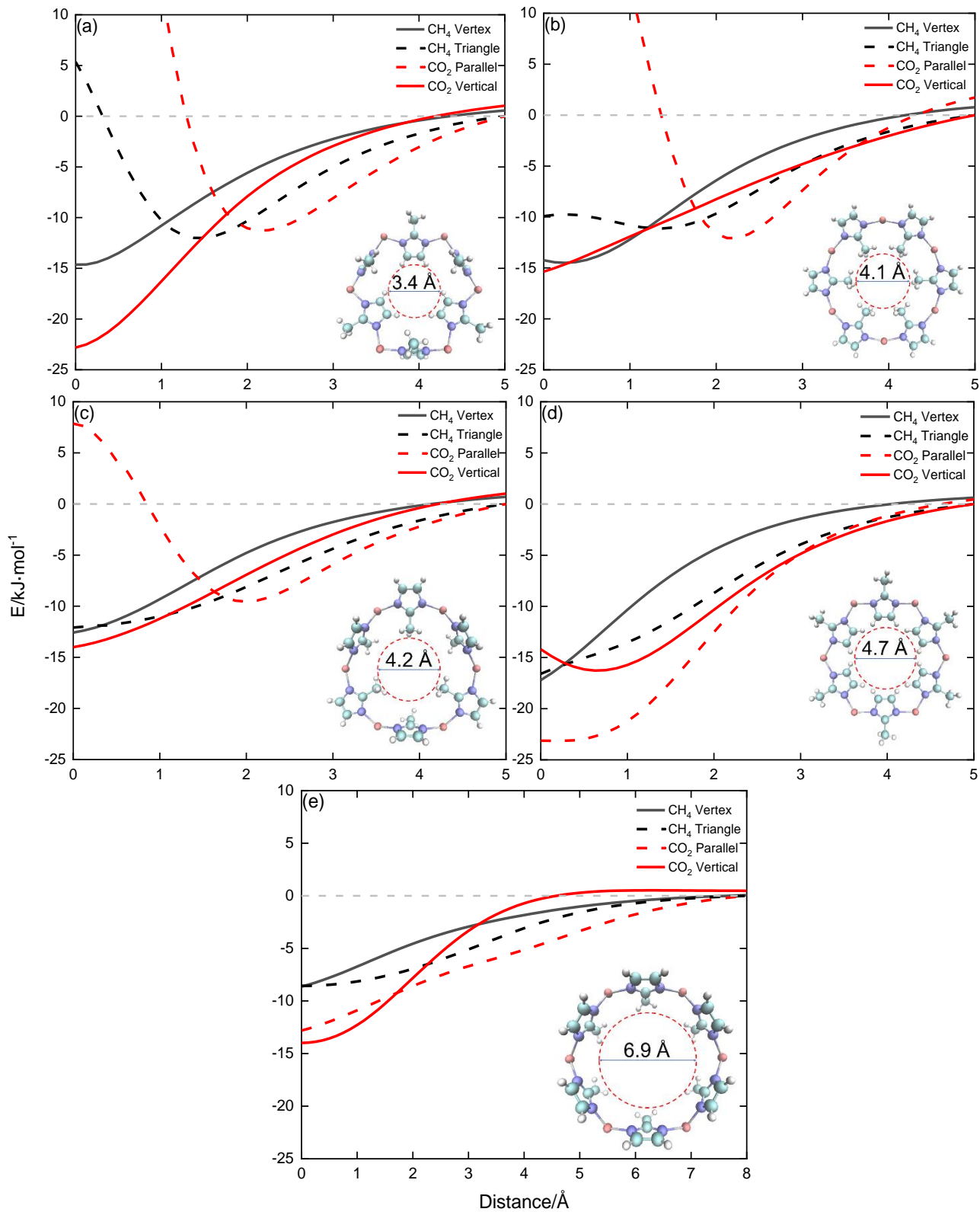

Figure 4. Potential energy curves for $CO_2$/$CH_4$ interacting with different ZIF-8 apertures. Distance is from the centre-of-mass of the $CO_2$/$CH_4$ molecule to the centre of the ZIF-8 aperture. The relative potential values are close to zero at the distance of (a)-(d): 5Å, (e): 8Å. Potential energies are given in kJ mol$^{-1}$.

The potential energy curves of $CO_2$/$CH_4$ molecules across different apertures of ZIF-8 are shown in Fig. 4. For $CO_2$, we find that the *'parallel'* orientation has much higher energy due to molecular interactions when the molecule is at the centre of the aperture. In Fig. 4(a), (b) and (c) the *'parallel'* orientations give positive high energy values (+88.1, +36.6, +7.9 kJ·mol$^{-1}$), while the *'vertical'* orientation gives negative energy values (-22.8, -15.4, -14.0 kJ·mol$^{-1}$) at the distance of 0.0 Å, indicating that the vertical orientation is more favourable for penetrating through these apertures (3.4 Å, 4.1 Å, 4.2 Å). Both Fig. 4 (d) and (e) show negative energy values for the two $CO_2$ orientations, indicating that both orientations could penetrate through these larger apertures (≥4.7 Å). This can be attributed to the electrostatic interactions between $CO_2$ and the aperture. The electrostatic potential (ESP) map of the apertures of 3.4 Å and 3.7 Å is presented in Fig. 5(a) and (b) respectively. The ESP map indicates the distribution of electrostatic potential with blue and red colours denoting positive and negative potentials, respectively. Compared to Fig. 5(a), 5(b) exhibits a stronger positive charge near the centre of the aperture. Therefore, the *'parallel'* orientation produces a stronger electrostatic attraction due to the positions of two negatively charged O atoms, resulting in a lower and negative energy.

For $CH_4$ molecules, the $CH_4$ *'triangle'* orientation gives positive energy values while the $CH_4$ *'vertex'* orientation gives negative energy values approaching the 3.4 Å ZIF-8 aperture (i.e. close to 0.0 Å), indicating the $CH_4$ *'vertex'* orientation being more favourable for penetrating the 3.4 Å aperture. For apertures which are bigger than 3.4 Å, both the *'triangle'* and *'vertex'* orientations of $CH_4$ give negative energy values approaching the aperture COM, although some variations in the energy curves are observed for different aperture size. A symmetry-adapted perturbation theory (SAPT) analysis[48] was performed to explain the irregular energy values (see Table S1). SAPT provides an approach of decomposing the interaction energies into several physically meaningful components, namely electrostatic, exchange, induction and dispersion contributions. Such method has been frequently applied to investigate the essence of interactions. In Fig. 4(a) and (b), the *'vertex'* position of $CH_4$ molecules gives lower energy values than the *'triangle'* within distance shorter than 1.2 Å and vice versa. In the rest of the aperture layouts, *'vertex'* always exhibits higher energy value indicating that the *'vertex'* layout is a more favourable $CH_4$ position. Fig. 4(a) presents a relatively larger energy deviation at distance of 0.0 Å for two different $CH_4$ positions (20.0 kJ·mol$^{-1}$) while Fig. 4(b) shows a smaller deviation of 4.3 kJ·mol$^{-1}$. Fig.4 (c), (d) and (e) give no obvious energy deviation at distance of 0.0 Å, with the single point energies being approximately -12.6, -17.2 and -8.6 kJ·mol$^{-1}$, respectively, for *'triangle'* and *'vertex'* positions. Note that Tian et al.[49] performed a similar approach of calculating potential energy curves of $CO_2$/$N_2$ molecules passing through porphyrin-based polymers. The single point energies for $CO_2$ molecules at the centre of the polymer lies within the range of ±20.0 kJ·mol$^{-1}$ depending on different porphyrin derivatives, which are on the same magnitude with our computed *'vertical'* energies.

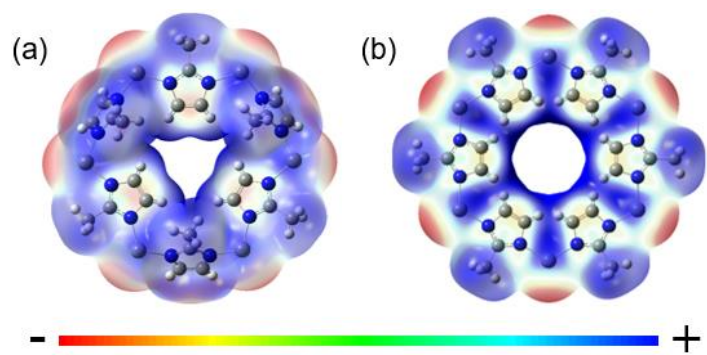

Figure 5. Electrostatic potential (ESP) map for aperture size = (a) 3.4 Å and (b) 4.7 Å. ESP maps are produced by GaussView 6[50]

Energy barrier refers to the energy values that gas molecules must overcome, in order to penetrate through the aperture. A large deviation in energy barriers usually result in a high selectivity of gas mixtures. The energy barrier values presented in Fig. 6 refer to the minimum energy that a single molecule ($CO_2$ or $CH_4$) needs to overcome when moving across the ZIF-8 aperture along the central line. It can be seen that for $CO_2$ molecules, the pristine and closed (4.7 Å) configuration present rather large energy barrier values of 22.8 and 23.1 kJ·mol$^{-1}$ correspondingly, suggesting that they are the most difficult aperture configuration for $CO_2$ molecules to pass through. These calculated energy barrier values are comparable to sap-3O porphyrin derivative reported by Tian et al.[49] (34 kJ·mol$^{-1}$). For *closed (4.1 Å)*, *semi-open* and *open* configurations, the $CO_2$ energy barriers are comparatively small (15.4, 14.0 and 10.6 kJ·mol$^{-1}$). For $CH_4$ molecules, the *closed (4.7 Å)* layout provides the highest energy barrier of 17.2 kJ·mol$^{-1}$, while *pristine* and *closed (4.1 Å)* layouts give nearly identical energy barrier values of 14.6 and 14.5 kJ·mol$^{-1}$, respectively. The *semi-open* configuration shows a moderately lower energy barrier of 12.6 kJ·mol$^{-1}$ with the *open* being the least difficult layout for $CH_4$ molecules to pass through. The deviation between $CO_2$/$CH_4$ energy barriers indicates the likely $CO_2$/$CH_4$ separation efficiency. It can be seen that the *pristine* aperture configuration (size = 3.4 Å) gives the best separation performance, as it shows the largest energy barrier value deviation (8.17 kJ·mol$^{-1}$). The *closed (4.1 Å)* and *semi-open* configuration exhibit poor separation performance due to negligible differences between the energy barrier values for $CO_2$ and $CH_4$ (0.9 versus 1.4 kJ·mol$^{-1}$). The difference of the separation efficiency between *pristine* and *closed (4.7 Å)* originates from $CH_4$ energy barriers. The higher $CH_4$ energy barrier value of the latter aperture configuration indicates a weaker $CH_4$ diffusivity. For the *open* configuration, the low energy barrier values as well as their deviation suggests a high molecule diffusivity but undesirable separation performance. Krokidas *et al.*'s work[51] in 2018 reported that the aperture size adjusts with the gas molecules which determines the gas diffusivity. The results here show that the aperture size is not a dominant factor for $CO_2$/$CH_4$ selectivity and no observable relations can be found between aperture size variation and separation efficiency. A larger size (open layout, 6.9Å) can have a better separation performance than smaller size apertures *(closed, 4.7Å)*. Instead, the aperture configuration plays a key role in the separation process.

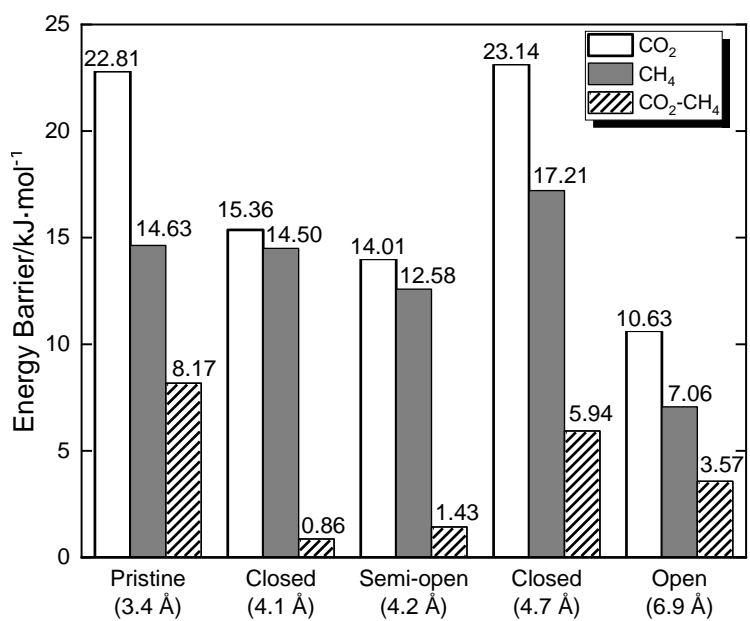

Figure 6. Energy barrier values of $CO_2$, $CH_4$ molecule passing through different ZIF-8 apertures. The difference between $CO_2$ and $CH_4$ is also given.

As concluded from DFT study, the *pristine* aperture configuration gives the best $CO_2/CH_4$ selectivity. In other words, less swinging during the gas mixture separation process is desired. Therefore, MD simulations were performed to investigate the variation of aperture configurations, namely geometry variation (GV). Fig. 7 (a) shows the GV of the two different systems of ZIF-8 and IL@ZIF-8 over 100 ns trajectories. The GV is evaluated by summing the distance between the current position and the original position ($t = 0$) for each atom. From the MD simulation results, we find that the *pristine* ZIF-8 exhibits larger GV than IL@ZIF-8 at 1bar, with fluctuation between 20-30 Å and 7-17 Å, respectively. Such results imply that the addition of [$C_4$mim][$Tf_2$N] reduced the GV (or swing effects) of ZIF-8, leading to a higher $CO_2/CH_4$ selectivity than pristine ZIF-8. When the pressure is increased to 20 bars, the GV is further diminished to 7-9 Å with smaller deviation. However, it should be pointed out that higher pressure does not ensure an enhanced $CO_2/CH_4$ separation performance, as evidenced from experimental data[27,28]. High pressure produces a low GV value, but also provides high momentum energy for gas molecules to overcome the energy barriers. Fig. 7(b) gives the record of structural variation during the MD simulations at 0 ns, 50 ns and 100 ns where the exemplary aperture is coloured in red. From the top and side views, it can be seen that the *pristine* ZIF-8 (without the addition of ILs) exhibited a structural change of larger extent, compared to the IL@ZIF-8 structures. This indicates that the addition of ILs could help stabilise the ZIF-8 structures, hence improve the $CO_2/CH_4$ separation performance.

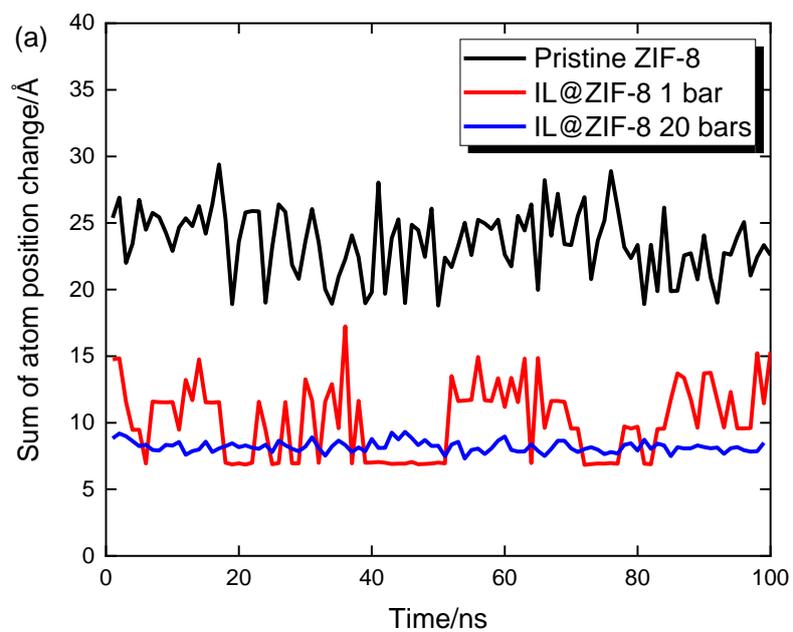

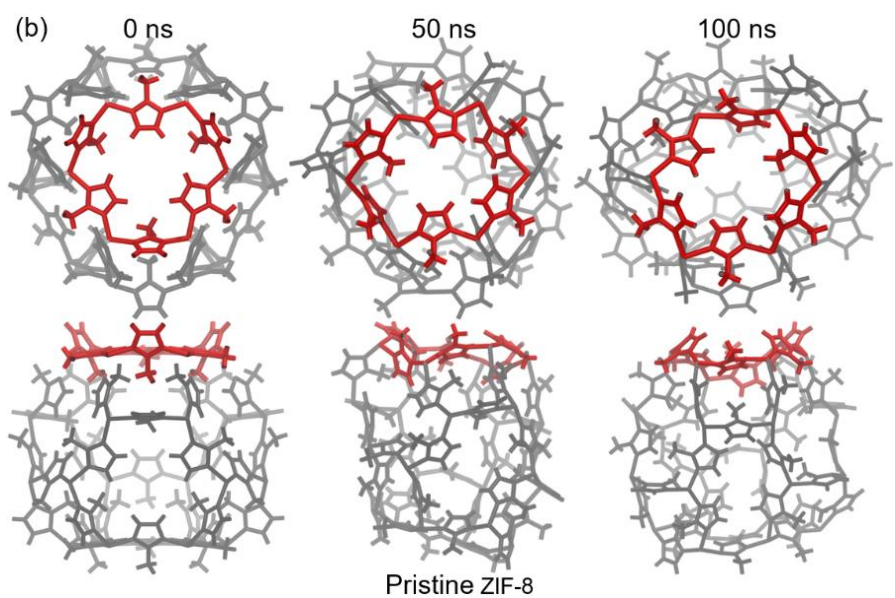

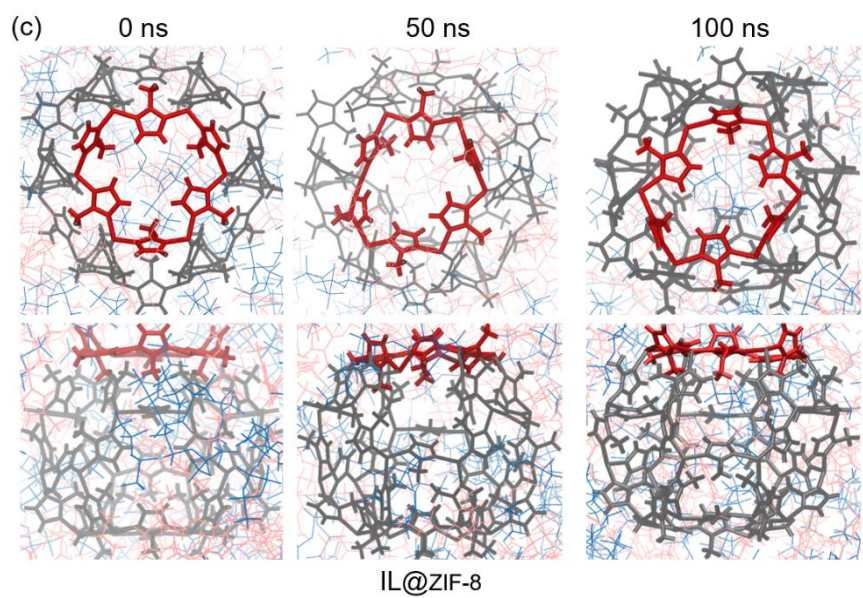

Figure 7. (a) MD simulation of geometry variation unit cell ZIF-8 and IL@ZIF-8. Top and side views of (b) pristine ZIF-8 and (c) IL@ZIF-8 at 0 ns, 50 ns and 100 ns MD simulations. The exemplary apertures are coloured in red.

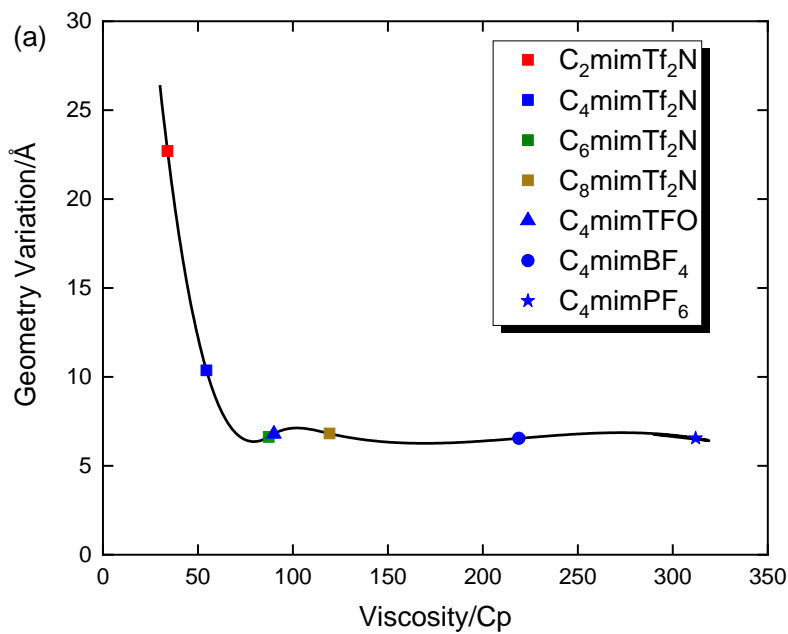
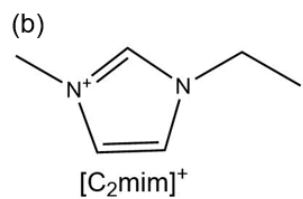
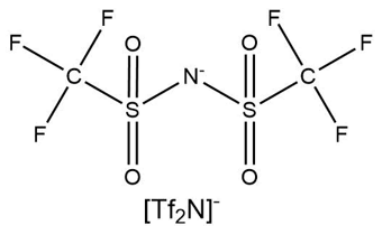
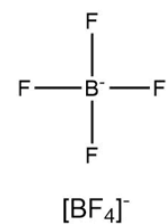
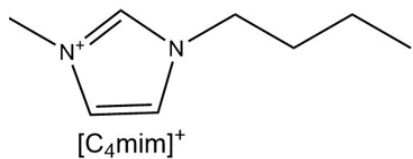
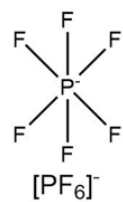
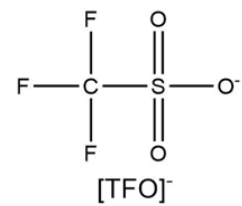
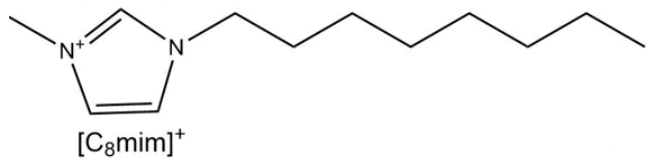

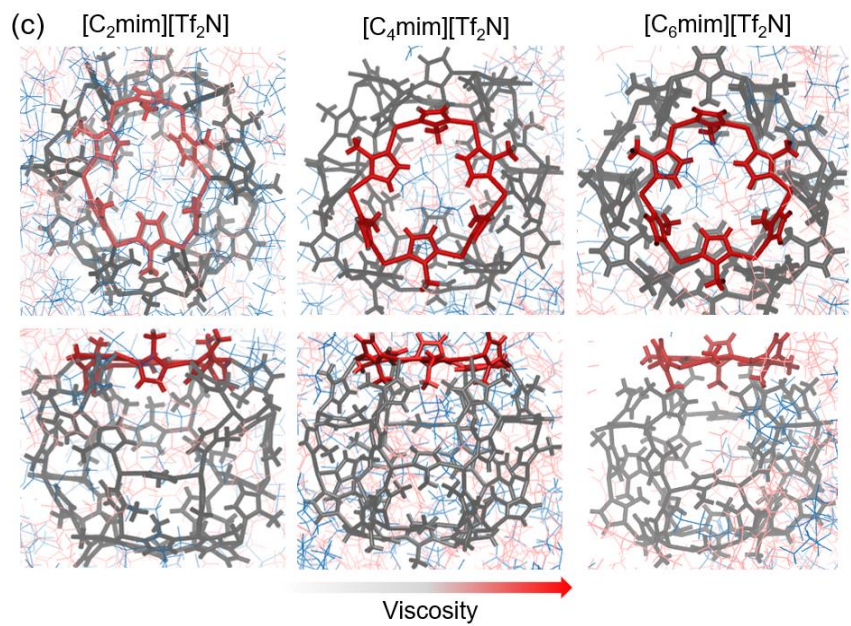

Figure 8. (a) MD simulation of Geometry variation of IL@ZIF-8 with different ILs: [C$_2$mim][Tf$_2$N], [C$_4$mim][Tf$_2$N], [C$_6$mim][Tf$_2$N], [C$_8$mim][Tf$_2$N], [C$_4$mim][TFO], [C$_4$mim][BF$_4$] and [C$_4$mim][PF$_6$]. Experimental viscosity values are taken from Zhang et al.[52] (b) Schematic representation of ILs. (c) Top and side views of [C$_2$mim][Tf$_2$N], [C$_4$mim][Tf$_2$N] and [C$_6$mim][Tf$_2$N] at 100ns MD simulations. The views of [C$_8$mim][Tf$_2$N], [C$_4$mim][TFO], [C$_4$mim][BF$_4$] and [C$_4$mim][PF$_6$] are given in Figure S1 in supporting information. Exemplary apertures are coloured in red. Cations and anions are coloured in light red and blue respectively.

## *3.2 Effects of ionic liquid on CO$_2$/CH$_4$ separation*

The effect of IL viscosities is further investigated by replacing [C$_4$mim][Tf$_2$N] with [C$_2$mim][Tf$_2$N], [C$_6$mim][Tf$_2$N], [C$_8$mim][Tf$_2$N], [C$_4$mim][TFO], [C$_4$mim][BF$_4$] and [C$_4$mim][PF$_6$], having different viscosities. The same MD simulations were performed for those ILs, and the mean GV value over 100 ns trajectories are given in Fig. 8(a). It can be seen that the GV value dramatically decreases from 22.7 Å to 10.4 Å when viscosity increases from 34 Cp to 54.5 Cp. The GV value continually declines to 6.6 Å when the viscosity is increased to 87.3 Cp. [C$_6$mim][Tf$_2$N] and [C$_4$mim][TFO] has very similar GV values (6.6 Å and 6.8 Å) due to very similar viscosity values (87.3 and 90 Cp). Note that further increase in viscosity above 87 Cp, no noticeable variation of GV value has been found, as depicted in Fig. 8 (a). The incorporation of [C$_6$mim][Tf$_2$N] in ZIF-8 shows a smaller GV, hence a higher selectivity than that with [C$_2$mim][Tf$_2$N], which is in agreement with previously reported experimental data[53]. Fig. 8(b) represents the structures of IL@ZIF-8 with different ILs at 100 ns of MD simulation. It can be observed that all ILs of viscosity larger than 87 Cp (ILs>87Cp) maintained their aperture configuration to their *pristine* configuration, with only very small variations. The GV value is mainly contributed from the rotation of imidazolates. [C$_4$mim][Tf$_2$N] exhibits larger extent of imidazolate rotation, resulting in a larger GV value of 10.4Å in comparison to ILs>87Cp. [C$_2$mim][Tf$_2$N] gives the largest GV value due to the variation of Zn-imidazolate-Zn angle. Overall, [C$_6$mim][Tf$_2$N] appears to be the best candidate for the IL@ZIF-8 system in terms of CO$_2$/CH$_4$ selectivity, separation performance and transfer properties.

## 4. Conclusion

In this work, DFT calculations were performed to evaluate the influence of aperture configurations of ZIF-8 on CO$_2$/CH$_4$ selectivity. Five different aperture configurations were chosen to cover possible energy barrier values. CO$_2$ and CH$_4$ were placed in different orientations and move across the aperture to obtain the potential energy curves. Results show that the *pristine* aperture configuration gives the best CO$_2$/CH$_4$ separation performance as it exhibits the largest energy barrier difference between CO$_2$ and CH$_4$. MD simulations were applied to investigate the effect of encaging IL [C$_4$mim][Tf$_2$N] in ZIF-8 (IL@ZIF-8) on CO$_2$/CH$_4$ separation. Simulated results indicate that the presence of IL enhanced the stability of *pristine* ZIF-8 aperture to retain its original configuration, leading to more efficient separation. MD simulations were conducted by replacing [C$_4$mim][Tf$_2$N] with ILs having different viscosities to explore the effect of viscosity on aperture configuration changes. We find that a more viscous IL gives a better separation performance, but no obvious change has been observed when viscosity value is greater than 87 Cp. In summary, the enhancement of CO$_2$/CH$_4$ selectivity was firstly explained from the aspect of aperture configuration variations based on DFT and MD calculations, to the best of our knowledge. Further experiments are needed to examine effects of aperture configurations on the gas mixture selectivity. Nevertheless, we believe that our derived conclusions would provide a new way to improve the gas selectivity of IL@MOFs composite materials.

**Conflicts of interest**

There are no conflicts to declare.

**Acknowledgments**

This work was financially supported by National Key R&D Program of China (2017YFB0603301), the National Natural Science Foundation of China (21978293), and CAS Pioneer Hundred Talents Program (L.L.).